\documentclass[conference]{IEEEtran}
\usepackage{amsmath,amsfonts}
\usepackage{algorithmic}
\usepackage{algorithm}
\usepackage{array}
\usepackage{subcaption}
\usepackage{textcomp}
\usepackage{stfloats}
\usepackage{url}
\usepackage[hidelinks]{hyperref}
\usepackage{verbatim}
\usepackage{graphicx}
\usepackage{cite}
\usepackage{glossaries}
\usepackage{siunitx}
\usepackage{flushend}
\usepackage{listings}
\usepackage{xcolor}
\usepackage{textcomp}

\definecolor{codegreen}{rgb}{0.25,0.47,0.47}
\definecolor{codegray}{rgb}{0.5,0.5,0.5}
\definecolor{codepurple}{rgb}{0.71,0.21,0.19}
\definecolor{backcolor}{rgb}{0.95,0.95,0.92}
\definecolor{emphcolor}{rgb}{0.08, 0.29, 0.62}

\lstdefinestyle{mystyle}{
    backgroundcolor=\color{white},   
    commentstyle=\color{codegreen},
    keywordstyle=\color{magenta},
    numberstyle=\tiny\color{codegray},
    stringstyle=\color{codepurple},
    basicstyle=\ttfamily\footnotesize,
    breakatwhitespace=false,         
    breaklines=true,                 
    captionpos=b,                    
    keepspaces=true,                 
    numbers=left,                    
    numbersep=5pt,                  
    showspaces=false,                
    showstringspaces=false,
    showtabs=false,                  
    tabsize=2,
    frame = single,
    emph={
        Constellation,
        BinarySource,
        LDPC5GEncoder,
        LDPC5GDecoder,
        Mapper,
        AWGN,
        Demapper,
        compute_ber,
        NeuralDemapper,
        BinaryCrossentropy,
        Dense,
        Layer,
        render,
        add,
        'scene.rt'},
    emphstyle=\color{emphcolor}
}
\newacronym{ACM}{ACM}{adaptive coding and modulation}
\newacronym{ADC}{ADC}{analog-to-digital conversion}
\newacronym{AGC}{AGC}{automatic gain control}
\newacronym{AOA}{AOA}{angle of arrival}
\newacronym{API}{API}{application programming interface}
\newacronym{AWGN}{AWGN}{additive white Gaussian noise}
\newacronym{AI}{AI}{artificial intelligence}
\newacronym{BER}{BER}{bit error rate}
\newacronym{BEC}{BEC}{binary erasure channel}
\newacronym{BLER}{BLER}{block error rate}
\newacronym{BP}{BP}{backpropagation}
\newacronym{BPA}{BP}{belief propagation}
\newacronym{BPTT}{BPTT}{back-propagation through time}
\newacronym{CFO}{CFO}{carrier frequency offset}
\newacronym{CN}{CN}{check node}
\newacronym{CIR}{CIR}{channel impulse response}
\newacronym{CNN}{CNN}{convolutional neural network}
\newacronym{CND}{CND}{check node decoder}
\newacronym{ConvNet}{ConvNet}{convolutional neural network}
\newacronym{CP}{CP}{cyclic prefix}
\newacronym{CRC}{CRC}{cyclic redundancy check}
\newacronym{CSI}{CSI}{channel state information}
\newacronym{DAC}{DAC}{digital-to-analog conversion}
\newacronym{DL}{DL}{deep learning}
\newacronym{DFT}{DFT}{discrete Fourier transform}
\newacronym{DOA}{DOA}{direction of arrival}
\newacronym{FFT}{FFT}{fast Fourier transform}
\newacronym{FEC}{FEC}{forward error correction}
\newacronym{GAN}{GAN}{generative adversarial network}
\newacronym{GRU}{GRU}{gated recurrent unit}
\newacronym{GPU}{GPU}{graphic processing unit}
\newacronym{GPS}{GPS}{global positioning system}
\newacronym{CPU}{CPU}{central processing unit}
\newacronym{HDPC}{HDPC}{high-density parity-check}
\newacronym{iid}{i.i.d.\@}{independent and identically distributed}
\newacronym{IFFT}{IFFT}{inverse fast Fourier transform}
\newacronym{ISAC}{ISAC}{integrated sensing and communications}
\newacronym{KL}{KL}{Kullback-Leibler}
\newacronym{LDPC}{LDPC}{low-density parity-check}
\newacronym{LLR}{LLR}{log-likelihood ratio}
\newacronym{LOS}{LoS}{line-of-sight}
\newacronym{LS}{LS}{least squares}
\newacronym{LSTM}{LSTM}{long short-term memory}
\newacronym{LMMSE}{LMMSE}{linear minimum mean squared error}
\newacronym{MIMO}{MIMO}{multiple-input multiple-output}
\newacronym{ML}{ML}{machine learning}
\newacronym{MLE}{MLE}{maximum likelihood estimation}
\newacronym{MLP}{MLP}{multilayer perceptron}
\newacronym{MRC}{MRC}{maximum ratio combining}
\newacronym{MSE}{MSE}{mean squared error}
\newacronym{MPA}{MPA}{message passing algorithm}
\newacronym{MMSE}{MMSE}{minimum mean squared error}
\newacronym{NLP}{NLP}{natural language processing}
\newacronym{NN}{NN}{neural network}
\newacronym{NLOS}{NLoS}{non-line of sight}
\newacronym{NeRF}{NeRF}{neural radiance field}
\newacronym{OFDM}{OFDM}{orthogonal frequency-division multiplexing}
\newacronym{OSS}{OSS}{open-source software}
\newacronym{pdf}{pdf}{probability density function}
\newacronym{pmf}{pmf}{probability mass function}
\newacronym{ReLU}{ReLU}{rectified linear unit}
\newglossaryentry{RIS}
{
  name={RIS},
  description={reconfigurable intelligent surface},
  first={\glsentrydesc{RIS} (\glsentrytext{RIS})},
  firstplural={\glsentrydesc{RIS}s (\glsentrytext{RIS})},
  plural={RIS}
}
\newacronym{RL}{RL}{reinforcement learning}
\newacronym{RNN}{RNN}{recurrent neural network}

\newacronym{RTN}{RTN}{radio transformer network}
\newacronym{SDR}{SDR}{software defined radio}
\newacronym{SER}{SER}{symbol error rate}
\newacronym{SFO}{SFO}{sampling frequency offset}
\newacronym{SNR}{SNR}{signal-to-noise ratio}
\newacronym{SIR}{SIR}{signal-to-interference ratio}
\newacronym{SINR}{SINR}{signal-to-interference-plus-noise ratio}
\newacronym{SGD}{SGD}{stochastic gradient descent}
\newacronym{SPA}{SPA}{sum product algorithm}
\newacronym{VN}{VN}{variable node}
\newacronym{VND}{VND}{variable node decoder}
\newacronym{SVM}{SVM}{support vector machine}
\newacronym{TDOA}{TDOA}{time difference of arrival}
\newacronym{TOA}{TOA}{time of arrival}
\newacronym{UAV}{UAV}{unmanned aerial vehicle}
\newacronym{wrt}{w.r.t.\@}{with respect to}
\newacronym{ZF}{ZF}{zero forcing}

\begin{document}

\title{Sionna RT: Differentiable Ray Tracing\\for Radio Propagation Modeling}

\author{
    Jakob Hoydis,    
    Fay\c{c}al A\"it Aoudia,
    Sebastian Cammerer,
    Merlin Nimier-David,\\
    Nikolaus Binder,    
    Guillermo Marcus,
    and Alexander Keller
}
\maketitle
\bstctlcite{IEEEexample:BSTcontrol}

\begin{abstract}
    Sionna\texttrademark{} is a GPU-accelerated open-source library for link-level simulations based on TensorFlow. Since release v0.14 it integrates a differentiable ray tracer (RT) for the simulation of radio wave propagation. This unique feature allows for the computation of gradients of the channel impulse response and other related quantities with respect to many system  and environment parameters, such as material properties, antenna patterns, array geometries, as well as transmitter and receiver orientations and positions. In this paper, we outline the key components of Sionna RT and showcase example applications such as learning radio materials and optimizing transmitter orientations by gradient descent. While classic ray tracing is a crucial tool for 6G research topics like reconfigurable intelligent surfaces, integrated sensing and communications, as well as user localization, differentiable ray tracing is a key enabler for many novel and exciting research directions, for example, digital twins.
\end{abstract}
\glsresetall

\section{Introduction}

Many 6G research topics require the simulation of specific radio environments by ray tracing. Examples are \gls{ISAC} \cite{sensing}, multi-modal sensing \cite{alkhateeb2022deepsense}, \glspl{RIS} \cite{ris}, radio-based localization \cite{studer2018channel}, \gls{ML}-based transceiver algorithms \cite{aiai}, as well as most of the use-cases of the recently started 3GPP study-item on AI/ML for the air interface \cite{sidai2021}. The reason for this is that a spatially consistent correspondence between a physical location in a scene and the \gls{CIR} is required which the widely used stochastic channel models such as \cite{3gpp38901} cannot provide. 
In addition, there is an increasing interest in ray tracing for the creation of digital twin networks \cite{alkhateeb2023real, lin20226g}. For these reasons, we have added a differentiable ray tracing module (RT) in release v0.14 of our open-source link-level simulator Sionna\texttrademark{} \cite{sionna} that we will introduce in this article.

Although ray tracing for radio propagation is a mature field \cite{iskander2015}, it has lately received renewed interest due to the development of \gls{RIS} \cite{9713744} and the potential of \gls{ML} techniques to improve accuracy and speed up computations \cite{deepray, ray-launching-wave-prop}. Several papers use \glspl{NN} for path loss prediction \cite{zhang2020cellular, yapar-u-net, 9954403, 9722715}, while \cite{winert} attempts to model ray-surface interactions by \glspl{NN}. The prevalence of \gls{ML} and \glspl{NN} is also apparent in the field of computer vision which is currently disrupted by neural rendering techniques such as \glspl{NeRF} \cite{muller2022instant}, while inverse rendering and differentiable ray tracing methods gain in popularity \cite{diffrt2018, Jakob2020DrJit, Mitsuba3}.

Driven by the aforementioned fusion of \gls{ML} and ray tracing, we have decided to develop Sionna RT as the world's first differentiable ray tracer for radio propagation  modeling. That means that functions of the generated field components, such as \glspl{CIR} and coverage maps, can be differentiated with respect to most parameters that are involved in their computation. These comprise the constituent material properties (conductivity $\sigma$ and relative permittivity $\varepsilon_r$), parameters of scattering functions, array orientations, positions, and geometries, as well as antenna patterns. In the future, we will include the configuration of \glspl{RIS} and multiple other ray-object interactions.

\section{Sionna RT}
Sionna RT is a ray tracing extension for radio propagation modeling that is built on top of Mitsuba~3 \cite{Mitsuba3} and TensorFlow \cite{tensorflow}. Like all of Sionna's components, it is differentiable, runs in a Jupyter notebook, is released under the Apache 2.0 license, and can be easily extended using high-level Python API calls that take full advantage of the available hardware, such as NVIDIA's GPUs. \Glspl{CIR} computed with Sionna RT can be immediately used for link-level simulations \textit{in-lieu} of a stochastic channel model.

\begin{figure}[t!]
    \centering
    \includegraphics[width=\columnwidth]{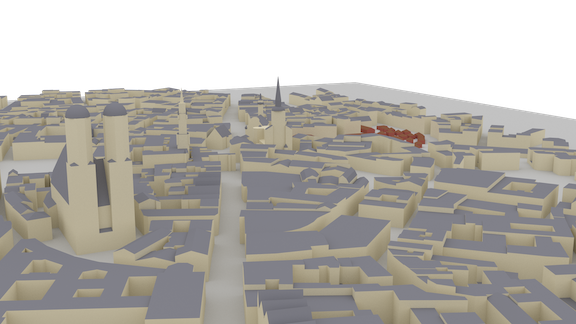}
    \caption{One of Sionna RT's example scenes. Data from \cite{osm}.}
    \label{fig:render_munich}
\end{figure}

Mitsuba~3 is a rendering system for forward and inverse light-transport simulation that makes use of the differentiable just-in-time compiler Dr.Jit \cite{Jakob2020DrJit}.
Sionna RT relies on Mitsuba~3 for the rendering and scene handling, e.g., its XML-file format, as well as the computation of ray intersections with scene primitives, i.e., triangles forming a mesh modeling a surface. The transformations of the polarized field components at each point of interaction between a ray and a scene object, e.g., reflections, are computed in TensorFlow, which is also used to combine the retained paths into (optionally) time-varying \glspl{CIR}. Thanks to TensorFlow's automatic gradient computation, \glspl{CIR} and functions thereof are differentiable with respect to many ray tracing parameters, including material properties, antenna patterns, orientations, and positions.

Scene files for Mitsuba~3 may be created, edited, and exported using the popular open-source 3D content creation suite Blender \cite{blender} and the Mitsuba-Blender add-on \cite{mitsuba-blender}. The Blender-OSM add-on \cite{blender-osm} allows one to rapidly create realistic scenes for almost any place in the world from OpenStreetMap \cite{osm}. In Sionna, scenes and radio propagation paths can be either rendered through the lens of configurable cameras via ray tracing or displayed with an integrated 3D viewer. For more detail on scene creation and rendering, we refer to Sionna's documentation\footnote{\url{https://nvlabs.github.io/sionna/api/rt.html}} and the video tutorial.\footnote{\url{https://youtu.be/7xHLDxUaQ7c}}

\begin{lstlisting}[language=Python, float, caption={Loading, rendering, and 3D preview of scenes in Sionna RT.}, label=lst:load_render, linewidth=245pt, belowskip=-0.8 \baselineskip]
import sionna
from sionna.rt import load_scene, Camera

# Load integrated scene
scene = load_scene(sionna.rt.scene.munich)

# Open 3D preview (only works in Jupyter notebook)
scene.preview()

# Render scene to file from preview viewpoint
scene.render_to_file(camera="preview",
                     filename="scene.png")

# Create new camera with different configuration
my_cam = Camera("my_cam", position=[0,-10,30])
scene.add(my_cam)

# Render scene with new camera
scene.render("my_cam")

# Render scene to file in high resolution
scene.render_to_file(camera="my_cam",
                     filename="scene2.png",
                     resolution=[1920, 1080])
\end{lstlisting}

Sionna RT allows for the definition of arbitrary radio materials which are characterized by their relative permittivity $\varepsilon_r$, conductivity $\sigma$, and several parameters related to scattering (scattering coefficient, pattern, cross polarization discrimination). Currently, only non-magnetic materials are supported, i.e., $\mu_r=1$. Radio materials can be defined by frequency-dependent functions, similar to the materials defined by the ITU in \cite{itu_p2040}, which are already integrated into Sionna.

An important design decision of Sionna RT is that all transmitters and all receivers in a scene have the same antenna array configurations. For example, all transmitters may be equipped with an $8\times 2$ dual-polarized array, while all receivers have a single cross-polarized antenna. Antenna arrays can be either explicitly modeled, i.e., paths are traced for every antenna element, or modeled synthetically after the ray tracing process by making a plane-wave assumption across the array. The former option is preferable for very large aperture arrays, where the plane-wave assumption does not hold. The latter option is significantly faster, especially in large setups. The antennas of an array can have arbitrary relative positions as well as user-defined and preconfigured radiation patterns.

\subsection{Code Examples}

Listing~\ref{lst:load_render} shows a code snippet that loads one of Sionna's integrated example scenes (line 5).
Once a \texttt{scene} is loaded, it can be visualized and explored within a Jupyter notebook using the \texttt{scene.preview()} function (line 8). A convenient feature is to render the current viewpoint of the preview into a file using \texttt{scene.render\_to\_file()} (line 11). The result is shown in Fig.~\ref{fig:render_munich}. Lines 15--16 show how a new \texttt{Camera} can be configured and added to the scene. By programmatically changing a camera's position and orientation, it is possible to generate images from different viewpoints, for example, as seen by a receiver navigating through the scene. As we will see later, also transmitter and receiver locations, coverage maps, and propagation paths can be easily visualized. 

\begin{figure}[t!]
    \vspace*{-10pt}
    \centering
    \includegraphics[width=\columnwidth]{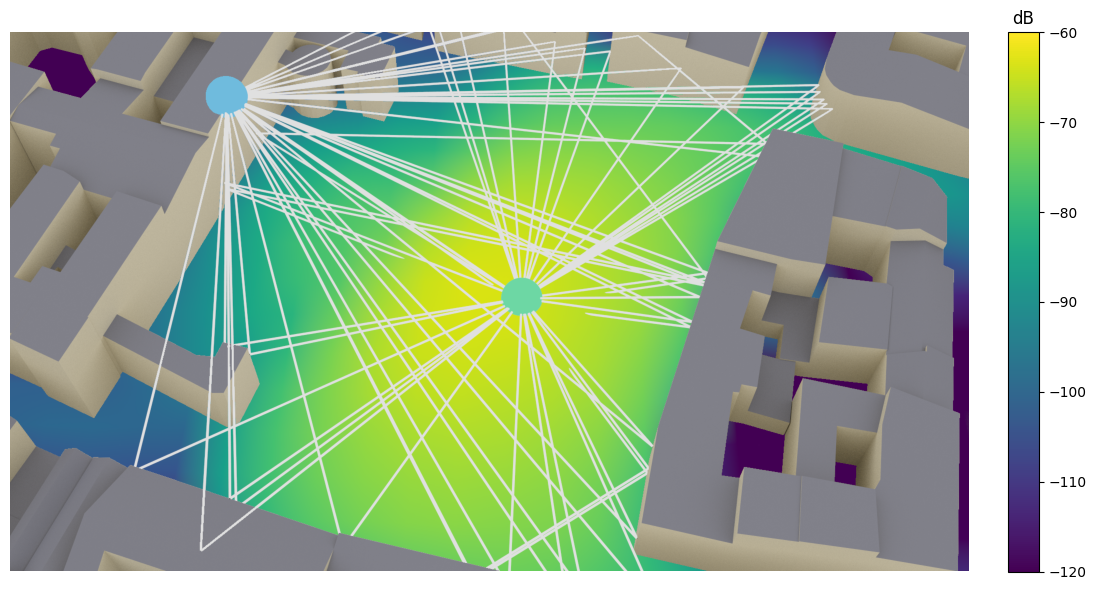}
    \caption{Visualization of the coverage map (path gain~[dB]) and ray-traced paths in Sionna RT from Listing~\ref{lst:tx_rx_paths}.}
    \label{fig:paths_vis}
\end{figure}

Listing~\ref{lst:tx_rx_paths} shows how to configure transmitters and receivers prior to computing propagation paths between them. All transmitters and all receivers are equipped with the same antenna arrays which are defined by the \texttt{scene} properties \texttt{scene.tx\_array} (line 30) and \texttt{scene.rx\_array} (line 38), respectively. Antenna arrays are composed of multiple identical antennas. Antennas can have custom or predefined patterns and are either single- or dual-polarized. One may add multiple transmitters and receivers to a scene. Besides a unique name, a position and orientation need to be provided. The latter  is defined by yaw, pitch, and roll angles. Like cameras, transmitters and receivers must be added to a scene (lines 46 and 55). The transmitter in our example is configured to be oriented towards the receiver (line 57). The function \texttt{scene.compute\_paths()} (line 61) executes the ray tracing process and computes propagation paths between all transmitters and receivers. The parameter \texttt{max\_depth} determines the maximum number of interactions between a ray and scene objects. For example, with a \texttt{max\_depth} of one, only \gls{LOS} paths are considered.
Different propagation phenomena such as diffraction and scattering can be additionally enabled.
When the property \texttt{scene.synthetic\_array} (line 60) is set to \texttt{False}, antenna arrays are explicitly modeled by finding paths between any pair of transmitting and receiving antennas in the scene. Otherwise, arrays are represented by a single antenna located at the center of the array.
Phase shifts related to the relative antenna positions will then be applied based on a plane-wave assumption when the \glspl{CIR} are computed. In line~65, a coverage map is generated which can be visualized together with the paths in the preview (line 69). Fig.~\ref{fig:paths_vis} shows a rendered image.

\begin{lstlisting}[language=Python, float, caption={Configuration of radio devices, path and coverage map computation, and visualization in Sionna RT.}, label=lst:tx_rx_paths, firstnumber=26, belowskip=-0.8 \baselineskip]
from sionna.rt import Transmitter, Receiver, \
                      PlanarArray

# Configure antenna array for all transmitters
scene.tx_array = PlanarArray(num_rows=8,
                             num_cols=2,
                             vertical_spacing=0.7,
                             horizontal_spacing=0.5,
                             pattern="tr38901",
                             polarization="VH")

# Configure antenna array for all receivers
scene.rx_array = PlanarArray(num_rows=1,
                             num_cols=1,
                             vertical_spacing=0.5,
                             horizontal_spacing=0.5,
                             pattern="dipole",
                             polarization="cross")

# Create transmitter
tx = Transmitter(name="tx",
                 position=[8.5,21,27],
                 orientation=[0,0,0])
scene.add(tx)

# Create a receiver
rx = Receiver(name="rx",
              position=[45,90,1.5],
              orientation=[0,0,0])
scene.add(rx)

tx.look_at(rx) # TX points towards RX

# Compute propagation paths
scene.synthetic_array = True
paths = scene.compute_paths(max_depth=3,
                            diffraction=True)

# Compute coverage map
cm = scene.coverage_map(max_depth=3,
                        diffraction=True)

# Visualize paths in the 3D preview
scene.preview(paths=paths, coverage_map=cm)
\end{lstlisting}

Once the paths are found, they can be transformed into \glspl{CIR} as shown in Listing~\ref{lst:cir}.
The class method \texttt{paths.apply\_doppler()} (line 2) can simulate the temporal evolution of the \glspl{CIR} based on arbitrary velocity vectors of all transmitters and receivers. \glspl{CIR} can then be obtained via \texttt{paths.cir()} (line 7) which allows deactivation of specific propagation phenomena to study their impact. The resulting \glspl{CIR} can then be used for link-level simulations in either time or frequency domains. Seamless end-to-end optimization of transceiver components~\cite{aiai} through the ray-traced channels is also possible.

\begin{lstlisting}[language=Python, float, caption={Generating channel impulse responses in Sionna RT.}, label=lst:cir, belowskip=-0.8 \baselineskip]
# Compute time-evolution depending on mobility
paths.apply_doppler(sampling_frequency=1e6,
                    num_time_steps=14,
                    tx_velocities = [3,0,0])
    
# Transform paths into channel impulse responses
a, tau = paths.cir(los=True,
                   reflection=True,
                   diffraction=False)
\end{lstlisting}

\subsection{Current Features \& Planned Extensions}
Sionna RT supports specular and diffuse reflections (i.e., scattering) as well as first-order diffraction. A detailed explanation of the involved computing steps is provided on the Sionna documentation website.\footnote{\url{https://nvlabs.github.io/sionna/em_primer.html}} Reflected paths are computed using the image method \cite{iskander_rt} by either exhaustive search, which does neither scale to large scenes nor to high reflection orders, or by a heuristic method. The heuristic method shoots a selected number of rays from the transmitter, whose directions are defined by a Fibonacci lattice on a sphere, which then propagate through the scene until the maximum number of reflections has been reached. For each resulting candidate path, i.e., an ordered list of scene primitives, the image method is applied to find specular paths that follow the identical list of primitives and arrive at the exact receiver location. Duplicate paths are removed and a possible \gls{LOS} path is added. 

For future releases, it is planned to add refraction as well as support for \glspl{RIS} while still ensuring differentiability of all involved processing steps.

\section{Example Applications} 
In this section, we will present two applications of differentiable ray tracing that are enabled by Sionna RT. The code to reproduce the results is available.\footnote{\url{https://github.com/NVlabs/diff-rt}}

\subsection{Learning Radio Materials}
Scene construction is an essential part of the ray tracing process. While scene geometries can be rather easily obtained today, e.g., from building databases such as OpenStreetMap \cite{osm}, there is no straightforward process to obtain the material properties. It is therefore important to develop techniques that automatically assign material properties, such as permittivity, conductivity, permeability, roughness, and scattering functions, to all objects in a scene.

Differentiable ray tracing allows us to optimize material properties based on data using gradient-based learning techniques. In this example application, we use Sionna RT to generate a data set of \glspl{CIR} from a scene with preconfigured radio materials. Pretending that the material properties
are unknown, we initialize them with default values and optimize them by gradient descent on the normalized mean squared error between the original channel frequency responses and those computed based on the \emph{trainable} materials. 

\begin{figure}[!t]
    \centering
    \includegraphics[width=0.95\columnwidth]{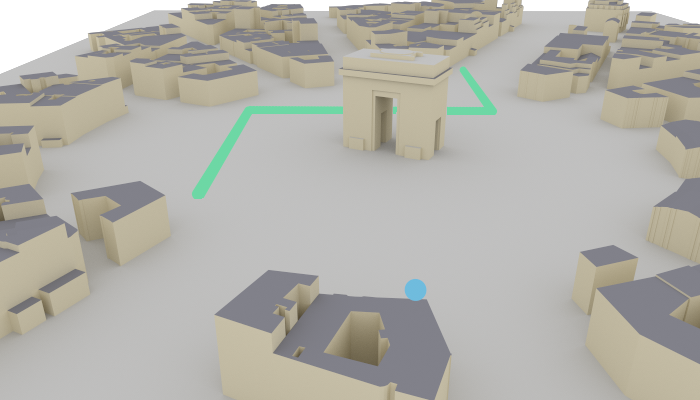}
    \caption{Data set generation for learning radio materials. The transmitter is located on a roof (blue) while \num{400} receivers (green) are located within the scene.}
    \label{fig:learning_materials}
\end{figure}

\begin{figure}[!t]
    \vspace{-8.5pt}
    \centering
    \begin{subfigure}{\columnwidth}
        \centering
        \includegraphics[width=0.9\columnwidth]{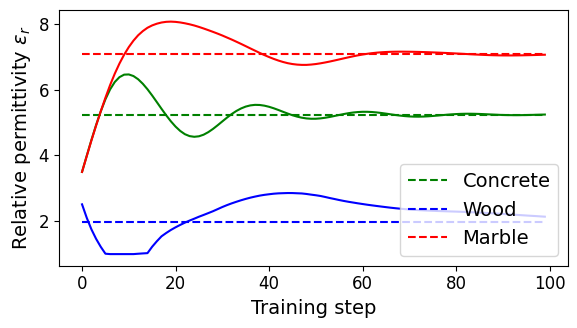}
    \end{subfigure}
    \begin{subfigure}{\columnwidth}
        \vspace{-4.5pt}
        \centering
        \includegraphics[width=0.9\columnwidth]{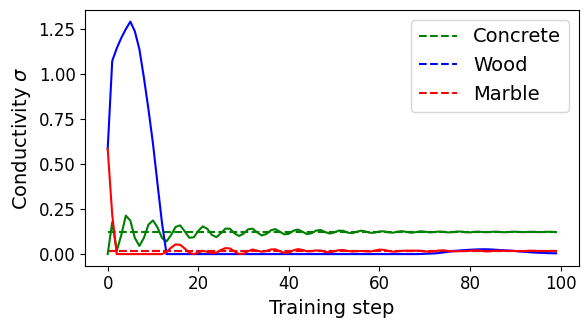}
    \end{subfigure}
    \caption{Training curves for radio materials.}
    \label{fig:training_curves}
\end{figure}

Fig.~\ref{fig:learning_materials} shows the locations of the transmitter and \num{400} receivers in the scene \texttt{sionna.rt.scene.etoile} for which \glspl{CIR} where generated. The original scene uses only four different materials: concrete for the streets, marble for the building walls, metal for the roofs, and wood for the ceiling underneath the Arc de Triomphe. We first generate \num{400} \glspl{CIR} that are transformed into the frequency domain assuming an \gls{OFDM} system with \num{128} subcarriers spaced \qty{30}{\kilo\hertz} apart. To start, the radio materials are replaced by trainable ones with an initial conductivity $\sigma$ of \qty{0.1}{\siemens\per\meter} and a relative permittivity $\varepsilon_r$ of \num{3.0}. Then, we iteratively compute channel frequency responses and update the material parameters using gradient descent until convergence. Fig.~\ref{fig:training_curves} shows the evolution of the learned radio materials during training. After around \num{100} iterations, the learned materials have converged to the targeted values indicated by dashed lines. Note that the convergence for wood is rather slow because only very few paths interact with objects made of it. We also do not show results for metal, as in our example there is no reflected path coming from a rooftop.

A few remarks are in order: First, our approach of training on channel frequency responses is unlikely to work well with measured data, as accurate phase information can rarely be predicted via ray tracing. A more practical approach is to train on coverage maps. This is an open field of research and many other options are possible.
Second, we have assumed that many objects in the scene share the same material. In practice, this may not be the case and one may need to assign a different material to each object or groups of objects instead. In turn, this will increase the training complexity. Third, only specular reflections are considered. Studying the impact of other propagation effects on the learning process, such as refraction, diffraction, and scattering, are interesting future investigations enabled by Sionna RT.

In practice, the data set for training may be obtained from geolocalized channel measurements. Note that the goal is not necessarily to obtain the correct material properties, but rather those that achieve the best prediction of the actually measured \glspl{CIR} in combination with the applied ray tracing algorithm. One may think about this as a sort of calibration of the ray tracer and environment model with respect to measurements.
Rather than optimizing material properties which are then used by physical models in the ray tracer to predict radio propagation, one may try to directly learn the interactions between rays and scene objects. A first exploration in this direction has been undertaken in \cite{winert}.

\subsection{Optimization of Transmitter Orientation}
Our second example application shows how differentiable ray tracing can be used to optimize the orientation of a transmit array in order to maximize the average received signal power in a specific region of the scene. The same approach may be used to optimize the joint orientation of multiple transmitters to maximize the weighted sum of \glspl{SINR} in different regions of the scene, or to optimize antenna geometries, precoding codebooks, configurations of \glspl{RIS}, as well as many other system parameters with respect to various metrics. It is also possible to solve inverse problems by gradient-based methods, such as determining a user's position based on measured \gls{CIR}.

\begin{figure}
    \centering
    \begin{subfigure}{\columnwidth}
        \centering
        \includegraphics[width=0.9\textwidth]{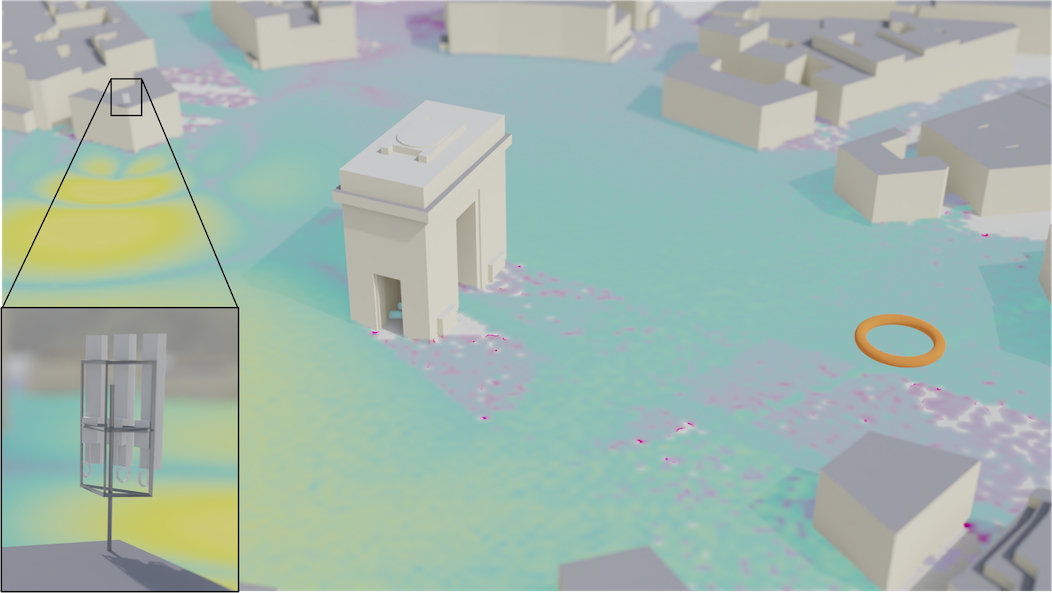}
        \caption{Before optimization}
        \label{fig:orientation_a}
    \end{subfigure}
    \begin{subfigure}{\columnwidth}
        \centering
        \includegraphics[width=0.9\textwidth]{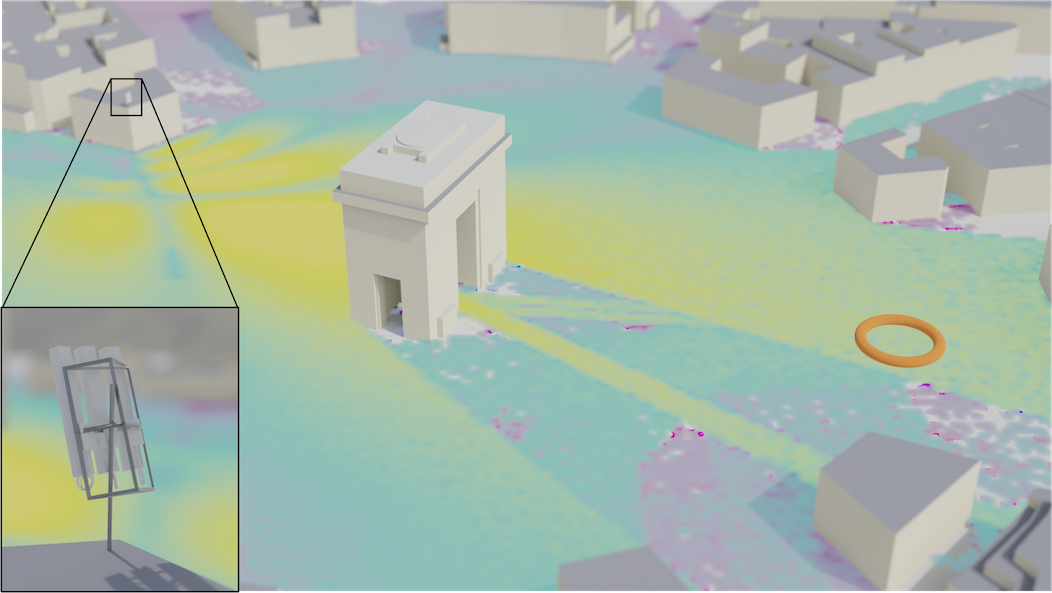}
        \caption{After optimization}
        \label{fig:orientation_b}
    \end{subfigure}
    \caption{Gradient-based optimization of the orientation of a transmitter (see the inset) with respect to the average received power within a small region of the scene (orange ring).}
    \label{fig:orientation}
\end{figure}

In this example, we place a transmitter on a building within the example scene \texttt{sionna.rt.scene.etoile} and compute a coverage map for a small region located behind the Arc de Triomphe (indicated by an orange ring). Then gradients of the average received power in this region with respect to the orientation of the transmitter are computed and used to optimize the latter via gradient ascent. Fig.~\ref{fig:orientation} shows the coverage map of the full scene before and after the optimization. Note that such optimization problems may become quickly non-convex and one would need to resort to other methods, such as Bayesian learning \cite{maggi2021bayesian}.

\section{Conclusions \& Future Directions}
Sionna RT is the world's first differentiable ray tracer for radio propagation modeling.
It enables many new avenues for research by integrating ray tracing, link-level simulations, and \gls{ML} into one open-source framework. One of our important ambitions is to make the creation of digital twin networks as simple as possible. We are confident that Sionna RT will enable the research to make this happen and are looking forward to contributions and feedback from the community.

\section*{Acknowledgements}
This work has received financial support from the European Union under Grant Agreement 101096379 (CENTRIC). Views and opinions expressed are however those of the author(s) only and do not necessarily reflect those of the European Union or the European Commission (granting authority). Neither the European Union nor the granting authority can be held responsible for them.

\bibliographystyle{IEEEtran}
\bibliography{IEEEabrv, bibliography}
\end{document}